\documentclass[pra,twocolumn,nofootinbib,notitlepage]{revtex4-1}

\usepackage{esdiff}

\usepackage{graphicx}
\usepackage{bm}
\usepackage{color}
\usepackage{amsmath, mathtools}
\usepackage{amssymb}
\usepackage{wasysym}
\usepackage{booktabs}
\usepackage{pdfsync}
\usepackage{verbatim}
\usepackage{latexsym}
\usepackage{dsfont}

\def\sec#1{\section{#1} }
\def\ssec#1{\subsection{#1} }

\def\({\left(}
\def\){\right)}
\def\[{\left[}
\def\]{\right]}

\def\a{\alpha}

\def\f#1#2{\frac{#1}{#2}}
\def\g{\gamma}

\def\de{\delta}

\def\k{\kappa}
\def\l{\lambda}

\def\p{\pi}
\def\r{\rho}

\def\th{\theta}

\def\ph{\phi}

\def\<{\langle}
\def\>{\rangle}

\begin{document}
\title{Robin boundary conditions are generic in quantum mechanics}

\author{Gwyneth Allwright}
\affiliation{Astrophysics, Cosmology and Gravity Centre,\\
Department of Mathematics and Applied Mathematics,\\
University of Cape Town\\
Rondebosch 7701, Cape Town, South Africa}

\author{David M. Jacobs}
\email{dmj15@case.edu}
\affiliation{Astrophysics, Cosmology and Gravity Centre,\\
Department of Mathematics and Applied Mathematics,\\
University of Cape Town\\
Rondebosch 7701, Cape Town, South Africa}

\begin{abstract}
Robin (or mixed) boundary conditions in quantum mechanics have received considerable attention in the last two decades, in particular, for applications to nanoscale systems. However, their utility has remained obscure to the larger physics community; in fact, one may find extant claims in the literature about the supposed naturalness of the specific Dirichlet (or vanishing) boundary condition. Here we demonstrate that not only are Dirichlet boundary conditions unnatural, but that Robin boundary conditions have utility in the long-wavelength approximation of short-ranged potentials.  For illustration, we consider a non-relativistic particle in one dimension under the influence of the multi-step and Morse potential. We derive the scattering and bound states for these potentials and determine the parameters that describe those states in an effective system, namely one in which there is a free particle with a boundary. This method appears to be generically applicable and also generalizable to many other real systems.
\end{abstract}

\maketitle

\sec{Introduction}
With an ever-growing effort going into nanoscale research, the ability to describe the features of small systems in a manner consistent with the laws of quantum physics remains a contemporary research topic. Self-adjoint extensions pertain to the allowable boundary conditions in quantum mechanics. They have been applied in pedagogical inquiries  \cite{Bonneau:1999zq, essin2006quantum}, but  also to more physically-motivated applications, such as characterizing point-like (or contact) interactions (e.g. \cite{jackiw1995diverse, Coutinho, Roy:2009vc, albeverio2012solvable}). A  complementary point of view is that an artificial boundary can be imposed in a model, for example, if there is a known limitation in experimental or model precision; boundary conditions are then used in place of the full (or UV-complete) model \cite{Jacobs:2015han}.

Nevertheless, it appears at this time that it is still widely unknown how and why boundary conditions are useful; furthermore, there is also a persistent misconception in the literature that the Robin (or mixed) boundary condition freedom, while of mathematical interest, is not physical.

Here we consider systems consisting of a non-relativistic particle confined to one dimension\footnote{This is also useful because wavefunctions on the half line indicate what happens for the radial part of higher-dimensional wavefunctions.} under the influence of a short-range potential; specifically, we are interested in the long-wavelength limit where the characteristic width(s) of the potential are much narrower than the de Broglie wavelength of the particles used to probe the system. Consequently, the waves interact with the potential in an approximately point-like manner. The aim here is to explicitly demonstrate that a particular one-parameter family of (Robin) boundary conditions for a wavefunction on the halfline can be used to approximate the physics of those short-range effects. Also, as we describe below, the Dirichlet condition is un-natural because it results from either a fine-tuning or from the extreme limiting cases of certain potentials, thereby contradicting extant claims in the literature, e.g. \cite{fulop2002classical, belchev2010robin, Kunstatter:2012np}.

First, consider a particle interacting with an infinite reflecting wall in one dimension. The potential for this system is 
\begin{equation}\label{free_particle_potential}
V(x)=
\begin{cases}
\infty,~~~~~&x<x_\text{b} \\
0,&x\geq x_\text{b}\,,
\end{cases}
\end{equation}
for some real $x_\text{b}$.  The standard shortcut to solve this rudimentary problem is to invoke continuity of the wave function; since the wavefunction must clearly vanish for $x<x_\text{b}$, it follows that
\begin{equation} \label{dirichlet}
\Psi(x_\text{b})=0\,.
\end{equation}
With that result, one then essentially disregards the portion of the $x$-axis for which $x<x_\text{b}$, thereby restricting the domain of the wavefunction to $x_\text{b} \leq x < \infty$, meaning there is effectively a Dirichlet boundary condition dictated by \eqref{dirichlet}. This restriction shouldn't be taken too seriously, however. The physical space clearly does not end at the boundary, but one need not consider what happens for $x<x_\text{b}$ since the wave function vanishes there.   Furthermore, although it may seem suspicious to require continuity of the wavefunction over a region where the potential is infinitely discontinuous, this standard method is supported by more rigorous analyses that consider regularized potentials that approach \eqref{free_particle_potential} as a particular limit \cite{fulop2002classical, belchev2010robin, Kunstatter:2012np}.

For a system with a boundary located at $x=x_\text{b}$, the most general boundary condition that preserves unitarity (or the Hermiticity of the Hamiltonian) is given by
\begin{equation}\label{robin_first}
\Psi(x_\text{b}) + L\, \Psi'(x_\text{b})=0\,,
\end{equation}
where the free parameter, $L$, has units of length and can take on any real value. Equation \eqref{robin_first} illustrates that there is an entire one-parameter family of Robin boundary conditions, from which Dirichlet is the special case in which $L = 0$. The conclusion by the authors of \cite{fulop2002classical, belchev2010robin, Kunstatter:2012np} is that $L=0$ follows from a perfect, infinitely reflecting wall. We do not disagree with that conclusion.

Our perspective, however, is that the discussion does not end there. All other \emph{non-zero} values of $L$ actually have a physical significance for all walls that are not perfect and infinite, i.e. any \emph{real} reflecting wall. In what follows below we will pay particular attention to Belchev and Walton's arguments \cite{belchev2010robin}. In their analysis, they place a potential well in front of large reflecting barrier, and take the limit in which the width of the well tends to zero, while the height of the barrier and depth of the well tends to infinity. This particular way of regularizing the potential allows one to recover a point-like interaction with the reflecting wall in which the Dirichlet boundary condition is the only natural choice, while all other boundary conditions require an extreme fine-tuning of the potential, making them physically unfeasible. These arguments, while mathematically consistent, are not well-motivated from a physical perspective.

\begin{center}
\begin{figure}[ht]
\includegraphics[width=0.48\textwidth]{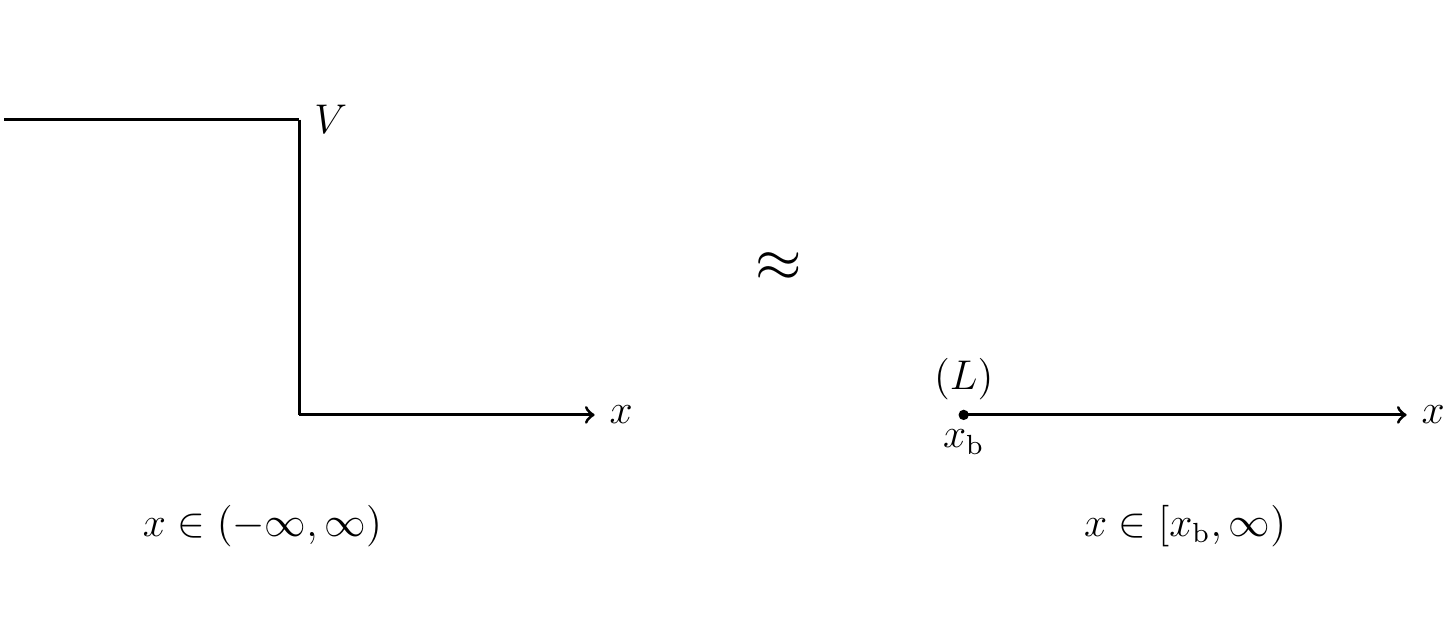}
\caption{Approximation of a system with an effective model, assuming that $E\ll V$, where $E$ is the energy of the scattered wave.}
\label{effective_model_pic}
\end{figure}
\end{center}

In the real world, the properties of potentials, such as the height of a potential barrier, cannot always be changed easily; more importantly, their characteristic quantities are never zero or infinitely large. On the other hand, one can adjust the momentum of the particles that are used to probe the system. With that in mind, a more fruitful way to approximate an interaction with a reflecting barrier is to analyze the system in the long-wavelength (or low-energy) limit. In this limit, if the scattering particle's momentum is much smaller than the inverse of any length scale of the potential, the interaction may be considered (at least perturbatively) point-like, as illustrated in Figure  \ref{effective_model_pic}. This works because the probability of the particle tunneling beyond the barrier is vanishingly small, or zero if the potential is truly infinite in extent. An initial investigation along these lines was first considered in \cite{Dasarathy:unpub}.

Here we reconsider the analysis of \cite{belchev2010robin} and illustrate how a system with zero potential, domain $[x_\text{b},\infty)$, and a boundary parameter, $L$, forms an effective low-energy model for a generic class of reflecting walls. Both $L$ and $x_\text{b}$ are free parameters, as far as the model is concerned, and in principle one may perform a set of experiments through which they may be measured. But, depending on the real underlying system, $L$ and $x_\text{b}$  will also have a correspondence with the parameters of a physical (UV-complete) model. In that case, the low-energy effective parameters can be derived by means of a matching calculation -- a measurable quantity, such as the phase shift, can be calculated with both the effective model and the long-wavelength limit of the underlying model. Ensuring agreement between both methods allows one to derive the effective parameters from first principles.

 In Section \ref{Effective_model_section} we explore the predictions of the effective model, then in Section \ref{Sec_multi-step} and \ref{Section_Morse} we consider the multi-step and Morse potentials, respectfully. In Section \ref{Conclusion} we discuss our results and open questions.

\sec{The Effective Model}\label{Effective_model_section}
Here a particle is constrained to the real line for $x\geq x_\text{b}$ with a boundary at $x=x_\text{b}$. The Hamiltonian $H$ for this system can be written as
\begin{equation}\label{free_hamiltonian}
H = -\f{1}{2m}\diff[2]{}{x}\,,
\end{equation}
where $m$ is the mass of the particle and $\hbar$ is set to unity (a convention we use throughout).  Several properties are required of observable operators that need not apply to operators that do not represent observables, specifically $H$ is self-adjoint (see e.g. \cite{Bonneau:1999zq}).  We will not bother to show it here, but it may be checked that \emph{all} wavefunctions must obey \eqref{robin_first} in order for the Hamiltonian to be a self-adjoint operator, and more physically, that unitarity is strictly enforced.

The time-independent Schr\"odinger equation corresponding to the Hamiltonian \eqref{free_hamiltonian} reads
\begin{equation}\label{halflineschro}
-\frac{1}{2m}\Psi''(x) = E\,\Psi(x)\,,
\end{equation}
where $E$ is the energy of the particle. Let $E\equiv k^2/2m>0$ so that, without loss of generality, the solutions to Equation \eqref{halflineschro} may be written
\begin{equation}\label{poshalf}
\Psi(x) \propto \sin{\(k(x-x_\text{b})+\theta\)}\,,
\end{equation}
where we will refer to $\th$ as the phase shift\footnote{Another perfectly valid form for the wavefunctions would be 
\begin{align}
\Psi(x)&\propto e^{-ik(x-x_\text{b})} + e^{+i\(k(x-x_\text{b})+2\de\)}\,,\notag
\end{align}
in which an incoming and scattered wave are explicitly represented. In this case the quantity $2\de$ is known as the \emph{total phase shift}. A Dirichlet ($L=0$) condition at $x=x_\text{b}$ would correspond to $2\de=\p$, so it may be useful to think of $2\th=2\de-\p$ as the \emph{additional} total phase shift, as compared to that of the canonical Dirichlet condition.}. This solution, taken with the Robin boundary conditions given in \eqref{robin_first} and imposed at $x=x_\text{b}$ imply
\begin{equation}\label{effective_tan_theta}
\tan{\theta} = -kL\,.
\end{equation}
Therefore, $L$ has the physical interpretation of the \emph{scattering length} of the effective model.

If $E<0$, then the normalizable solutions to Equation \eqref{halflineschro}, satisfying the boundary condition in \eqref{robin_first}, are of the form
\begin{equation}
\Psi(x) \propto e^{-\f{x-x_\text{b}}{L}}\,,
\end{equation}
and the single bound-state energy $E_B$ is
\begin{equation}\label{eb}
E_B = -\f{1}{2mL^2}\,.
\end{equation}
A bound state is only possible for a positive decay length, i.e. $L>0$; however, it is not guaranteed. Although the effective theory may predict a bound state, one must keep in mind that there will be cutoff length scale, $\l_\text{cut}$, below which the theory cannot be trusted; for the scattering states this corresponds to $k\l_\text{cut}\ll1$. If $L \gg \l_\text{cut}$ then the bound state should exist, but  if $L \lesssim \l_\text{cut}$ the bound state is likely spurious. Examples are given in the following sections.

A simple time-of-flight experiment could be used to determine $x_\text{b}$. Classically, if a particle with momentum $k$ in the direction of the boundary begins and ends its motion at $x=x_0$, then the classical time of flight is simply
\begin{equation}\label{effective_TOF_classical}
\Delta t_\text{cl}=\f{2m (x_0 - x_\text{b})}{k}\,.
\end{equation}
As addressed in \cite{belchev2010robin}, an additional (Wigner) time delay, $\de t_\text{delay}$ would be observed due to quantum effects:
\begin{equation}\label{Wigner_delay}
\de t_\text{delay} = -\f{2m L}{k\(1+k^2 L^2\)}\,.
\end{equation}
However, equation \eqref{effective_TOF_classical} is sufficient to determine $x_\text{b}$. When considering other potentials in Sections \ref{Sec_multi-step} and \ref{Section_Morse}, we will calculate the associated classical times-of-flight as
\begin{equation}\label{TOF_classical}
\Delta t_\text{cl} = 2\int_{x_\text{cl}}^{x_0}~\f{dx}{\sqrt{\f{2}{m}\(E-V(x)\)  }  }\,,
\end{equation}
where $x_\text{cl}$ is defined by $E=V(x_\text{cl})$.  By matching  \eqref{effective_TOF_classical} to \eqref{TOF_classical} we will then extract $x_\text{b}$.

In the next two sections we consider two reflecting-wall models and derive the value of $L$ and $x_\text{b}$ that correspond to them, justifying our claim that the interactions can be modeled by means of a Robin boundary condition.

\begin{figure}
\includegraphics[width=0.5\textwidth]{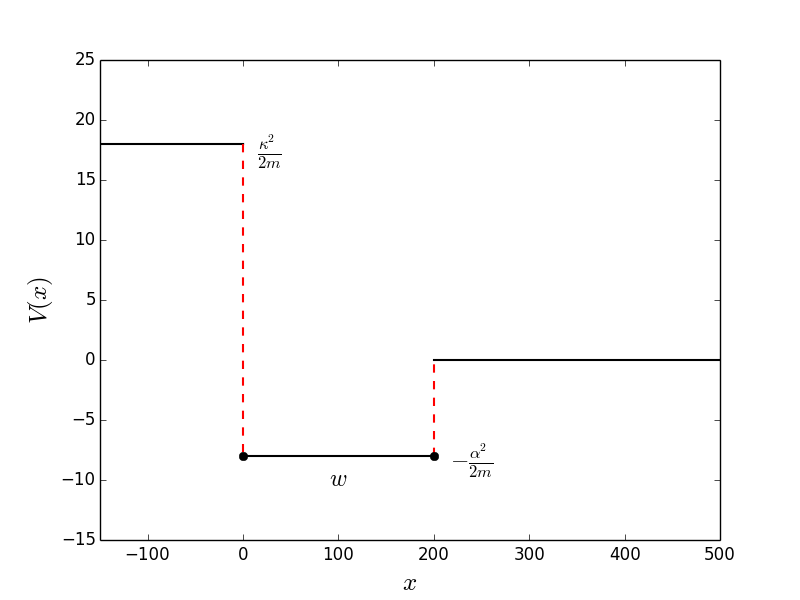}
\caption{Plot of the multi-step potential with $m=1$, $w=200$, $\alpha=4$ and $\kappa=6$.}
\label{multi_potential_plot}
\end{figure}

\sec{The Multi-Step Potential}\label{Sec_multi-step}
For convenience we define the multi-step potential as
\begin{equation}\label{multi_pot}
V(x)=
\begin{cases}
\f{\k^2}{2m},~~~~~&x<0\\
-\f{\a^2}{2m},&0\leq x \leq w\\ 
0,&w<x\,,
\end{cases}
\end{equation}
where $\a$, $\k$ and $w$ are real positive constants; see Figure \ref{multi_potential_plot} for an illustration. One recovers the infinite reflecting wall \eqref{free_particle_potential} as a limit of this potential, e.g., as  $\k\to\infty$ and $w\to0$.

The classical time-of-flight here, given by equation \eqref{TOF_classical}, is
\begin{equation}
\Delta t_\text{cl}= \f{2m}{k}\(x_0 - w\)\,,
\end{equation}
to lowest order in $k/\a$, and by comparison with equation \eqref{effective_TOF_classical} we learn that $x_\text{b}=w$.

\ssec{Scattering States}
The time-independent Schr\"odinger equation for a single particle is given by
\begin{equation}\label{tise_general}
-\frac{1}{2m}\Psi''(x) + V(x)\Psi = E\,\Psi(x)\,,
\end{equation}
where $V(x)$ is the potential given in \eqref{multi_pot} and $E$ is the energy of the particle. Let $0<E<{\kappa^2}/{2m}$ and define $E\equiv k^2/2m$. Then the solutions to Equation \eqref{tise_general} with Equation \eqref{multi_pot} as the potential are of the form
\begin{equation}\label{scatsols}
\Psi(x) \propto
\begin{cases}
\exp\({p x}\),~~~~~&x<0\\
\sin\({p' x+\chi}\),&0\leq x \leq w\\
\sin\({k(x-w)+\Theta}\),&x > w\,,
\end{cases}
\end{equation}
where 
\begin{equation}
p \equiv \sqrt{\kappa^2-k^2}\,\,\,\,\,\,\,\,\,\,\,\,\,\,\,\text{and}\,\,\,\,\,\,\,\,\,\,\,\,\,\,\,p' \equiv \sqrt{k^2+\alpha^2}\,.
\end{equation}
Since $x_\text{b}=w$, a comparison with equation \eqref{poshalf} allows us to immediately identify $\Theta=\th$, where $\th$ is the phase shift in the effective model. We would therefore like to find the low-energy limit of $\th(k)$ using the full model, and use it to determine the low-energy parameter, $L$. It can be computed analytically by  simply matching the wavefunction and its first derivative at $x=0$ and $x=w$, giving
\begin{equation}\label{full_thingey}
\tan{\th} = \frac{{k}Q - {p'}\tan{kw}}{p'+ {k}Q\tan{kw}}\,,
\end{equation}
where
\begin{equation}
Q \equiv \frac{p' + p\tan{p'w}}{p-p'\tan{p'w}}\,.
\end{equation}
For the matching calculation we desire the long-wavelength limits, 
\begin{equation}
kw\ll1\,,\,\,\,\,\,\,\,\,\,\,k\ll \k\,\,\,\,\,\,\,\,\,\, \text{and} \,\,\,\,\,\,\,\,\,\,k\ll \a\,,
\end{equation}
in which case
\begin{equation}
Q\simeq \frac{\alpha + \kappa\tan{\alpha w}}{{\kappa} - {\alpha}\tan{\alpha w}}\,,
\end{equation}
and hence,
\begin{equation}\label{scat_phase_approx}
\tan{\th} \simeq - k\(w + \frac{1 + \frac{\kappa }{\alpha}\tan{\alpha w}}{\alpha  \tan{\alpha w}-\kappa }\).
\end{equation}
A final comparison of equation \eqref{effective_tan_theta} with \eqref{scat_phase_approx} therefore implies
\begin{equation}\label{MS_L}
L = w + \frac{1 + \frac{\kappa }{\alpha}\tan{\alpha w}}{\alpha  \tan{\alpha w}-\kappa }\,.
\end{equation}
This is the value of the boundary parameter that reproduces the phase $\theta$ for the full system in the low-energy limit. As advertised, it would require an extreme fine-tuning of the underlying model parameters for $L=0$ (or $L\to \pm\infty$, for that matter) to result.

\ssec{Bound States}
Let ${-\alpha^2}/{2m}<E<0$ and define $E\equiv -q^2/2m$. The solutions to Equation \eqref{tise_general} with \eqref{multi_pot} as the potential are of the form
\begin{equation}
\Psi(x) \propto
\begin{cases}
\exp\({\sqrt{\kappa^2+q^2}x}\),~~~~~&x<0\\
\sin\({\sqrt{\alpha^2-q^2}\,x+\beta}\),&0\leq x \leq w\\
\exp\({-q\(x-w\)}\)\,,&w<x
\end{cases}
\end{equation}
Applying the continuity of the wave function and its derivative at $x=0$ and $x=w$ gives
\begin{equation}\label{boundstateenergy}
\tan\({\sqrt{\alpha^2-q^2}\,w}\) = \f{\sqrt{\alpha^2-q^2}\(q + \sqrt{\kappa^2+q^2}\)}{\alpha^2-q^2-q\sqrt{\kappa^2+q^2}}\,,
\end{equation}
from which the bound-state energies may be obtained; generally, this has to be done numerically since this equation is transcendental.

Using the effective parameter found in \eqref{MS_L}, we would predict a single bound state within the effective model with momentum scale
\begin{equation}\label{qeff_Multi-step}
q_\text{eff}=L^{-1}= \(w + \frac{1 + \frac{\kappa }{\alpha}\tan{\alpha w}}{\alpha  \tan{\alpha w}-\kappa }\)^{-1}\,.
\end{equation}
One might therefore expect to find a correspondence between one of the bound states of the full system, determined by \eqref{boundstateenergy}, and this bound state; however, that does not generically happen. One can view this from two perspectives. First of all, the bound state energies are not experimentally tunable like they are for scattering states, so there exists no low-energy limit to speak of. Second, for the effective model to work, i.e. a point-like interaction to be a good approximation, we would expect that the decay length of the wavefunction, $q^{-1}\gg w$. This could only occur for the least bound state, and would only occur if there is tuning between $\a, \k$ and $w$.  For illustration, consider the simplified case wherein $\k\gg\a$ and $\k\gg w^{-1}$; therefore, 
\begin{equation}
L\simeq w -\a^{-1}\tan{\a w}\,.
\end{equation}
It must be that $L/w\gg1$ for the predicted bound state to actually exist; if there is a physical mechanism that would cause the product $\a\, w$ to be just a bit larger than an odd multiple of $\p/2$, i.e. if
\begin{equation}
\a\, w = \(\f{(2j+1)\pi}{2}+\de\)\,,
\end{equation}
where $j$ is a non-negative integer and $\de\ll1$, then
\begin{equation}
q_\text{eff}=L^{-1}\simeq \f{\(2j+1\)\pi}{2w}\de\,,
\end{equation}
and this would be approximately consistent with a root of \eqref{boundstateenergy}.

Such tuning does occur for certain underlying physical models; a nice example is a nuclear binding model in which a 3-dimensional potential well has a depth and width that are both characterized by the same fundamental (nuclear) scale. This is presumably why a boundary condition approach lends itself well to describing the binding of light nuclei \cite{Jacobs:2015han}.  While there is a class of systems for which this effective method describes actual bound states, it does not generically work unless $L$ is much greater than the cutoff scale, $\l_\text{cut}$; this could be determined empirically from scattering experiments, alone, but for this multi-step potential we have derived $\l_\text{cut}=\max{(\k^{-1}, \a^{-1},w)}$. When $L\lesssim\l_\text{cut}$ the bound state must be discarded as spurious.

\begin{figure}
\includegraphics[width=0.5\textwidth]{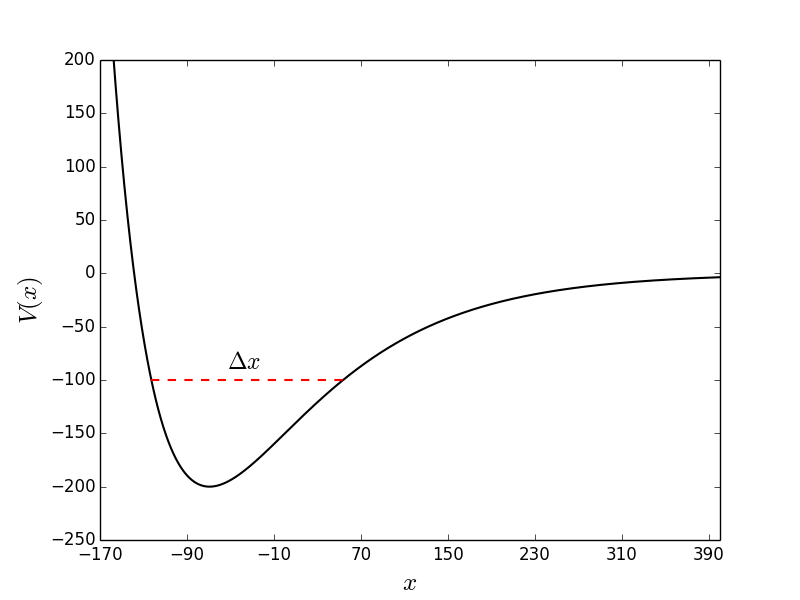}
\caption{Plot of the Morse potential $V(x)$ with $m=1$, $\a=0.01$, $b=4$ and $\kappa=10$. }
\label{morse_plot}
\end{figure}

\sec{The Morse Potential}\label{Section_Morse}
Following Reference \cite{belchev2010robin}, we parametrize the Morse potential as\footnote{Compared to  \cite{belchev2010robin}, we have set the coefficient $b=1$; this corresponds to a redefinition of $x$ and $\k$ and does not reduce the generality our analysis.}
\begin{equation}\label{morse_def}
V(x)=\f{\k^2}{2m}\(e^{-2\a x}- e^{-\a x}\)
\end{equation}
where $\alpha$ and $\kappa$ are real constants with dimensions of momentum.  As shown in Figure \ref{morse_plot}, the Morse potential grows exponentially large for negative $x$, while it decays exponentially for positive $x$,  providing a smoothed version of the multi-step potential. 

The salient features of this potential are its minimum of
\begin{equation}
V(x_\text{min})=-\f{\kappa^2}{8m}\,,
\end{equation}
where $x_\text{min}=\alpha^{-1}\ln{2}$, and a well-like structure around $x_\text{min}$ with a characteristic width, $\Delta x\simeq\a^{-1}$. In the limit $\a\to\infty$ it follows that $\Delta x\to 0$, and the infinite reflecting wall \eqref{free_particle_potential} is recovered.

To derive $x_\text{b}$, the classical time-of-flight is calculated here using the potential given in equation \eqref{morse_def} in equation \eqref{TOF_classical}; it is analytical tractable, but cumbersome, so we will not display the full results. Noting that the classical turning point is
\begin{align}\label{xcl}
x_\text{cl}
&\approx 0\,,
\end{align}
for low energies, the salient feature of the time-of-flight calculation is that
\begin{align}
\Delta t &\simeq \f{2m}{k}\(x_0 + \f{2}{\a} \ln{\(\f{k}{\k}\)}\)\notag \\
& \approx \f{2m}{k}x_0
\end{align}
which assumes $k\ll\a$ and $k\ll\k$, but that $x_0$ is macroscopically large so that $k\, e^{\a x_0} \gg \k$. Comparison with equation \eqref{effective_TOF_classical} indicates that $x_\text{b}=0$.

\ssec{Scattering States}

The Schr\"odinger equation for this system is
\begin{equation}\label{first_schro}
-\f{1}{2m}\Psi''(x)+ \f{\k^2}{2m}\(e^{-2\a x}- e^{-\a x}\)\Psi(x)=E\, \Psi(x)\,.
\end{equation}
Setting $E\equiv k^2/(2m)$ and defining $z\equiv e^{-\a x}\,$, $\Psi(z)\equiv z^{i k/\a}  e^{-\k z/\a} g(z)$, and $\f{2\k}{\a} z\equiv y$, one obtains the following equation for $g(y)$:
\begin{equation}\label{CHE}
y g''(y) + \(B-y\)g'(y)    -Ag(y)=0\,,
\end{equation}
where
\begin{equation}
B\equiv\f{2ik}{\a}+1
\end{equation}
and
\begin{equation}
A\equiv\f{1}{2}\(\f{2ik}{\a}+1 - \f{\k}{\a}\)\,.
\end{equation}
Equation \eqref{CHE} is known as the confluent hypergeometric equation (CHE), or Kummer's differential equation, 
 whose solution may be written as the generic linear combination
\begin{equation}
g(y)\!=\!d_1M(A,B,y)+ d_2y^{1-B} M(A-B+1,2-B,y)\,,
\end{equation}
where
\begin{equation}\label{kummer}
M(A,B,y)=\sum_{n=0}^\infty  \f{1}{n!}\f{A^{(n)}}{B^{(n)}} y^{n}
\end{equation}
is known as Kummer's function, $d_1$ and $d_2$ are numerical constants, and the notation $A^{(n)}=A(A+1)\dots(A+n-1)$ denotes a rising factorial. Hence, for a particular $k$, the full solution to the time-independent Schr\"odinger equation with the Morse potential is given by
\begin{multline}\label{general_psi_solution}
\Psi(x)\sim e^{-i k x} e^{-\k z/\a} \[d_1M\(A,B,\f{2\k z}{\a}\)\right.\\
~~~~~~\left.+ d_2\(\f{2\k z}{\a}\)^{1-B} M\(A-B+1,2-B,\f{2\k z}{\a}\)\]\,.
\end{multline}

This solution must be normalizeable in the $x\to -\infty$ ($z\to \infty$) limit. According to \cite[p. 504]{abramowitz+stegun}, 
\begin{equation} \label{mlim}
\lim_{y\to\infty}M(A,B,y)\sim \f{\Gamma(B)}{\Gamma(A)}e^y y^{A-B}+\mathcal{O}\(\left |y\right |^{-1}\)\,,
\end{equation}
which generically makes \eqref{general_psi_solution} non-normalizeable since, for $x<0$, it blows up as $\exp{\(\k/\a\, e^{-\a x}\)}$. However, there exists a special combination of $d_{1,2}$ that kills off this divergent behavior. Defining
\begin{equation}\label{nice}
U(A,B,y)\!\equiv\! d_1 M(\!A,\!B,y\!) + d_2 y^{1-B} M(\!A-B+1,\!2-B,y\!)\,,
\end{equation}
then equation \eqref{mlim} indicates
\begin{align}\label{nice2}
\lim_{y\to\infty}U(A,B,y) &= d_1 \f{\Gamma(B)}{\Gamma(A)}e^y y^{A-B} \notag \\
&+ d_2  \f{\Gamma(2-B)}{\Gamma(A-B+1)}e^y y^{A-B} + \mathcal{O}\(\left |y\right |^{-1}\)\,.
\end{align}
We require that the $e^y$ terms in Equation \eqref{nice2} vanish by an appropriate choice of $d_1$ and $d_2$. Up to an overall normalization constant, the unique choice is
\begin{align}\label{tricomi}
U(A,B,y)&=  \f{\Gamma{(1-B)}}{\Gamma{(A-B+1)}}  M(A,B,y)  \notag \\
&+ \f{\Gamma(B-1)}{\Gamma(A)} y^{1-B} M(A-B+1,2-B,y)\,,
\end{align}
also known as the Tricomi confluent hypergeometric function.

Finally, the physical solution to the time-independent Schr\"odinger equation with the Morse potential is given by
\begin{equation}\label{general_psi}
\Psi(x) \sim e^{-i k x} e^{-\k z/\a} U\(A,B,2\k z/\a\)\,,
\end{equation}
up to a normalization constant. This is consistent with Belchev and Walton's result, equation (27) of \cite{belchev2010robin}; however, this derivation did not rely on any spurious requirements, such as ``reality of the wavefunction".

To determine the phase shift we take the $x\to\infty$ ($z\to 0$) limit of the wave function and, because the potential vanishes there, we expect to find that the wave function approaches that of a free particle,
\begin{align}
\lim_{x\to\infty}\Psi(x)
&\equiv e^{-ikx} - e^{+i\(kx+2\ph\)}\label{phi_def}\,.
\end{align}
Since $x_\text{b}=0$, comparison with equation \eqref{poshalf} allows us to immediately identify $\phi=\th$, where $\th$ is the phase shift in the effective model. We would therefore like to find the low-energy limit of $\th(k)$ using the full model, and use it to determine the low-energy parameter, $L$. It is straightforward to show that
\begin{align}\label{psi_limit}
\lim_{x\to\infty}\Psi(x)&\sim \(\f{\Gamma(1-B)}{\Gamma(A-B+1)}\)e^{-ikx} \notag \\
&~~~~ + \(\f{\Gamma(B-1)}{\Gamma(A)}\(\f{2\k}{\a}\)^{-2ik/\a} \)e^{ikx}\,,
\end{align}
up to a normalization constant and therefore
\begin{align}\label{shift2}
e^{2i\th}
&=-\f{\Gamma\(-\f{2ik}{\a}\)}{\Gamma\(+\f{2ik}{\a}\)}\f{\Gamma\(\f{\a - \k -2ik}{2\a}\)}{\Gamma\(\f{\a - \k +2ik}{2\a}\)}\(\f{2\k}{\a}\)^{-2ik/\a}\,.
\end{align}

The next step is to expand Equation \eqref{shift2} in terms of small $k$. In order to do this, we will use Euler's infinite product formula, which is 
\begin{equation}\label{infprod}
\Gamma(t) = \f{1}{t}e^{-\g t}\prod_{n=1}^\infty \(1+\f{t}{n}\)^{-1} e^{t/n}\,,
\end{equation} 
where $\g=0.5772\dots$ is the Euler-Mascheroni constant. We are then able to put equation \eqref{shift2} in the more useful form:
\begin{align}\label{shiftprod}
e^{2i\th}=&\exp{\[-\f{2ik}{\a}\(\ln\({\f{2\k}{\a}}\)+\g\)\]} \(\f{\a - \k +2ik}{\a - \k -2ik}\)  \notag \\
&\times\prod_{n=1}^\infty\f{(1-\f{2ik}{\a n})(1+\f{\a - \k +2ik}{2\a n})}{(1+\f{2ik}{\a n})(1+\f{\a - \k -2ik}{2\a n})}\exp\({\f{2ik}{\a\,n}}\)\,.
\end{align}

Next, we take the argument of the right-hand-side of Equation \eqref{shiftprod} to find an expression for $\th$. Since the argument of a product of complex numbers is simply the sum of the arguments of each individual complex number, we can convert the infinite product in Equation \eqref{shiftprod} into an infinite sum. Additionally, we use the fact that for a complex number, $z=\r\, e^{i\chi}$, the argument $\chi$ can be determined by means of
\begin{equation}
\arg\({z}\)=\chi=\arctan\[{    \f{\text{Im} (z)}{\text{Re} (z)} }\]\,,
\end{equation}
and, as a corollary, $\arg\({z/\bar{z}}\)=2\arg\({z}\)$, where $\bar{z}$ denotes the complex conjugate of $z$. Therefore it follows that
\begin{multline}\label{almost_0}
\th=  -\f{k}{\a}\[\ln\({\f{2\k}{\a}}\)+\g\] + \arctan{\f{2k}{\a -\k}}~~~\\
+ \sum_{n=1}^\infty\[\arctan{\(  \f{2k}{\a(2n+1)-\k} \)}-\arctan{\(\f{2k}{\a n}\)}+\f{k}{\a n}\]\,.
\end{multline}

Since we seek an expression that is valid in the long wavelength limit, we then expand Equation \eqref{almost_0} to first order in $k$, yielding
\begin{align}\label{almost}
\th&=  -\f{k}{\a}\[\ln\({\f{2\k}{\a}}\) + \g  - \f{1}{1/2-\k/(2\a)} \right. \notag \\
 &~~~~~~~~~~+\left. \sum_{n=1}^\infty\(\f{1}{n} - \f{1}{n+1/2-\k/(2\a)}\) \] +\mathcal{O}\(k^3\) \notag \\
&\simeq  -\f{k}{\a}\[\ln\({\f{2\k}{\a}}\) + \g + \sum_{j=0}^\infty\(\f{1}{j+1} - \f{1}{j+\f{1}{2}-\f{\k}{2\a}}\) \]\,.
\end{align}
The second line is a more useful form because of the identity \cite{abramowitz+stegun}
\begin{equation}\label{id}
\psi(z)=-\g +\sum_{m=0}^\infty\(\f{1}{m+1}-\f{1}{m+z}\)\,,
\end{equation}
where $\psi(x)$ is the digamma function, and $z\neq0,-1,-2,\dots$. Therefore, Equation \eqref{almost} becomes 
\begin{equation}\label{finalshift}
\th=-\f{k}{\a}\[2\g + \ln\({\f{2\k}{\a}}\) + \psi\(\f{1}{2}-\f{\k}{2\a}\)\]\,,
\end{equation}
to first order in $k$. Comparing to the small $k$ ($kL\ll1$) limit of equation \eqref{effective_tan_theta} means that 
\begin{equation}\label{Morse_L}
L = \f{1}{\a} \[2\g + \ln\({\f{2\k}{\a}}\) + \psi\(\f{1}{2}-\f{\k}{2\a}\)\]\,.
\end{equation}
The fact that $L\propto \a^{-1}$ is not surprising since that is the characteristic width of the potential well.

\ssec{Bound States}

For bound states, we define
\begin{equation}
E\equiv -\f{q^2}{2m}\,.
\end{equation}
Solving the Schr\"odinger equation follows as above, with the substitutions $z\equiv e^{-\a x}\,$ and $\f{2\k}{\a} z\equiv y$. Defining
\begin{equation}
\tilde{B}\equiv\f{2q}{\a}+1
\end{equation}
and
\begin{equation}
\tilde{A}\equiv \f{2q+\a-\k}{2\a}\,,
\end{equation}
the generic solution to the Schr\"odinger equation is the linear combination
\begin{align}
\Psi(x)\propto &e^{-q x} e^{-\k z/\a} \[d_1M\(\tilde{A},\tilde{B},\f{2\k z}{\a}\) \right.\notag \\
&+\left. d_2\(\f{2\k z}{\a}\)^{1-\tilde{B}} M\(\tilde{A}-\tilde{B}+1,2-\tilde{B},\f{2\k z}{\a}\)\]\,.
\end{align}
However, just as above, normalizeability in the $x\to -\infty$ ($z\to \infty$) limit requires a precise combination of $d_{1,2}$ so that 
\begin{equation}
\Psi(x) \sim e^{-q x} e^{-\k z/\a} U\(\tilde{A},\tilde{B},\f{2\k z}{\a}\)\,,
\end{equation}
We also require that $\Psi(x)$ vanishes as $x\to \infty$ ($z\to 0$), meaning that
\begin{align}
\lim_{x\to\infty}\Psi(x)  &\sim e^{-q x} e^{-\k z/\a}\notag\\
&~~\times \(    \(\f{2\k z}{\a}\)^{1-\tilde{B}} \f{\Gamma(\tilde{B}-1)}{\Gamma(\tilde{A})}  + \f{\Gamma(1-\tilde{B})}{\Gamma(\tilde{A}-\tilde{B}+1)}   \) \notag\\
&\sim e^{+q x} \(\f{2\kappa}{\alpha}\)^{-2q/\alpha}\f{\Gamma(\tilde{B}-1)}{\Gamma(\tilde{A})}
\end{align}
must vanish. These states will not be normalizeable unless this divergent behavior can be avoided; this occurs if (and only if) $A$ is equal to a non-positive integer, or likewise
\begin{equation}
q(n)=\f{\k - (2n+1)\a}{2}\,,
\end{equation}
where $n$ is a non-negative integer. Since our convention is that $q >0$, there are only bound states if $\a < \k$. There are a total of $n_\text{max} +1$ states, where the maximum quantum number,
\begin{equation}
n_\text{max} = \text{floor}\[\f{1}{2}\(\f{\k}{\a}-1\)\]\,.
\end{equation}

The ``least bound" state corresponding to $q(n_\text{max})$ generically will have a decay length of ${\cal O}(\a^{-1})$, in which case the state is confined to a region too small to be described by the effective system. However, it would be captured by the effective system if $q(n_\text{max}) \ll \a$, which occurs for the specific relation between $\k$ and $\a$,
\begin{equation}
\k = (2 n_\text{max}+1+\de)\,\a\,,
\end{equation}
for $0<\de \ll 1$. In that particular case one finds
\begin{equation}
q_\text{max} = \f{\de}{2} \a \,.
\end{equation}
The effective model could then be used to predict the bound state from the scattering information via equation \eqref{Morse_L},
\begin{align}
L^{-1} &=  \f{\a}{2\g + \ln\({4n_\text{max}+2+2\de}\) + \psi\(-\f{2 n_\text{max}+\de}{2}\)}\notag \\ 
& \simeq \f{\de}{2} \a \,,
\end{align}
which follows because $\psi\(- n_\text{max} + \f{\de}{2}\)\sim 2/\de+{\cal O}(\de^0)$.

\sec{Conclusion}\label{Conclusion}

Here we have demonstrated that Robin boundary conditions can be used to approximate the physics of short-range interactions if they are probed at long wavelength. Two distinct UV-complete systems were considered: one with the multi-step potential, and another with the smoother and more physically realistic Morse potential. Both were compared to an effective system consisting of a free particle moving on a half-line with a boundary at $x=x_\text{b}$ and a corresponding Robin boundary condition specified by the boundary parameter, $L$.  In the low-energy limit, both UV-complete systems could be described by the effective system to good approximation. These conclusions appear to be quite general and also contradict earlier claims in the literature about how the Dirichlet boundary condition is natural. Beyond its pedagogical interest, we hope this perspective on the use of boundary conditions may be of some practical use, especially in the areas of nanoscale research.

Open questions remain, for example, whether or not this type of approach will be useful for describing real systems consisting of time-dependent quantum wells that are too difficult to solve analytically (e.g. \cite{kashcheyevs2010universal}). It would also be interesting to see if it is possible to extend this analysis by promoting the boundary parameter to be a function of the particle energy, i.e. $L\to L(E)$, and seek a series expansion in $E$. This would seem to make a stronger connection between this approach and that of effective range theory  \cite{Bethe:1949yr}. A superficial assessment suggests that this is problematic because every wavefunction would then satisfy a different boundary condition dictated by \eqref{robin_first}; one consequence would be a lack of strict unitarity. However, if unitarity is preserved in a time-averaged sense, this may still prove to be useful. This is an open question and the subject of followup work.

\section*{Acknowledgements}
We would like to acknowledge Harsh Mathur and Kate Brown for stimulated interested in, and many conversations about, the use of boundary conditions in quantum mechanics. We also thank Glenn Starkman, Andrew Tolley, and Will Horowitz for useful conversations. One of the authors (D.M.J.) would like to acknowledge support from the Claude Leon Foundation.

\bibliographystyle{apsrev}
\bibliography{RobinBCsAreGeneric}
\end{document}